\documentclass{aastex}
\usepackage{graphicx}

\usepackage{emulateapj5}
\usepackage{times}

\makeatletter

\newenvironment{inlinefigure}{%
\def\@captype{figure}%
\noindent\begin{minipage}{0.999\linewidth}\begin{center}}
{\end{center}\end{minipage}\smallskip}
\makeatother

\def\keV{ke\kern-0.05emV}
\newcommand{\chandra}{\emph{Chandra}}

\def\gsim{\mathrel{\hbox{\rlap{\hbox{\lower4pt\hbox{$\sim$}}}\hbox{$>$}}}}
\def\lsim{\mathrel{\hbox{\rlap{\hbox{\lower4pt\hbox{$\sim$}}}\hbox{$<$}}}}

\def\chandra    {{\em Chandra}\/}
\def\rxj1720    {{ RXJ1720.1+2638}\/}

\def\degd       {$^{\circ}\!$}
\def\second     {{\prime\prime}}
\def\ms1455     {{MS 1455.0+2232}}

\begin{document}

\submitted{2002, ApJL, 569, 31}

\title{\chandra ~ Observation of a $300$~kpc Hydrodynamic  
Instability in the Intergalactic Medium of the Merging Cluster of Galaxies A3667}

\author{Pasquale\ Mazzotta\altaffilmark{1,2}, Roberto\ Fusco-Femiano
\altaffilmark{3},
 Alexey\ Vikhlinin \altaffilmark{1}}
\altaffiltext{1}{Harvard-Smithsonian Center for Astrophysics, 60 Garden St.,
Cambridge, MA 02138; mazzotta@cfa.harvard.edu}
\altaffiltext{2}{Department of Physics, University of Durham,
South Road, Durham DH1 3LE}
\altaffiltext{3}{Istituto Astrofisica Spaziale, Area CNR Tor Vergata,
via del Fosso del Cavaliere, 00133 Roma (Italy)}

\begin{abstract}
  
  We present results from the combination of two \chandra ~ pointings of the
  central region of the cluster of galaxies A3667. From the data analysis of
  the first pointing Vikhlinin et al. reported the discovery of a
  prominent cold front which is interpreted as the boundary of a cool gas
  cloud moving through the hotter ambient gas. 
Vikhlinin et al. discussed the
    role of the magnetic fields in maintaining the apparent dynamical
    stability of the cold front over a wide sector at the forward edge of the
  moving cloud
 and suppressing
 transport processes across the front.  
  In this Letter, we identify two new 
  features in the X-ray image of A3667: i) a 300 kpc arc-like
  filamentary X-ray excess extending from the cold gas cloud border into the
  hotter ambient gas; ii) a similar arc-like filamentary X-ray depression
  that develops inside the gas cloud.  Both features are located beyond the
  sector identified by the cold front and are oriented in a direction
  perpendicular to the direction of motion. The temperature map suggests
  that the temperature of the filamentary excess is consistent with that
  inside the gas cloud while the temperature of the depression is consistent
  with that of the ambient gas.  We suggest that the observed features
  represent the first evidence for the development of a large scale
  hydrodynamic instability in the cluster atmosphere resulting from a major
  merger.  
  This result confirms previous claims for the
  presence of a moving  cold gas cloud into the 
  hotter ambient gas.
  Moreover it shows that, although the gas mixing
  is suppressed at the leading edge of the subcluster due to 
  its magnetic structure, strong turbulent mixing 
  occurs at larger angles to the direction of motion.
  We show that this mixing process may favor the deposition of a 
  nonnegligible quantity of thermal energy right in the cluster 
  center,  affecting the development of the central cooling flow.

\end{abstract}


\section{Introduction}

The central region of A3667, a nearby, hot merging cluster (Markevitch,
  Sarazin, \& Vikhlinin 1999), was observed for the first time by \chandra
~ in Sept 1999.  
The analysis of this
  first observation by Vikhlinin, Markevitch \& Murray (2001a,b) reveals
the presence of a prominent 500~kpc-long density
  discontinuity (``cold front'') in the cluster atmosphere. 
Vikhlinin et al.\ show that: 
i) the density discontinuity is the
contact surface between the moving 
cloud of 4~keV gas
and the hotter ambient gas; ii) the speed of the moving
cloud is slightly supersonic.
  
An interesting aspect raised by 
Vikhlinin et al. is that 
the cold front
must develop hydrodynamical instabilities. 
These instabilities would destroy the front on very short time
scales. 
However, the \chandra ~ image shows that
the front is stable in a wide, $\varphi =\pm 30^\circ$ sector (where $\varphi$ 
is the angle with respect to the direction of motion). Moreover,
the front width is smaller than the Coulomb mean free path, which indicates
that the transport and mixing processes are suppressed (see also Ettori \&
Fabian 2000). To reconcile this observational evidence with the
expected hydrodynamical instability of the cold front, 
Vikhlinin et al. (2001b) put forward the
following scenario. As the cloud moves through the ambient gas, the
magnetic field lines, which are frozen in the intracluster medium, are
stretched by the tangential plasma motions. This results in the formation of
a layer with the ordered magnetic field parallel to the front surface. The
magnetic field intensity in such a layer is sufficient to suppress the
hydrodynamical instabilities within the $30^\circ$ sector, and to prevent gas
mixing and transport processes.

\begin{inlinefigure}
\centerline{\includegraphics[width=0.95\linewidth]{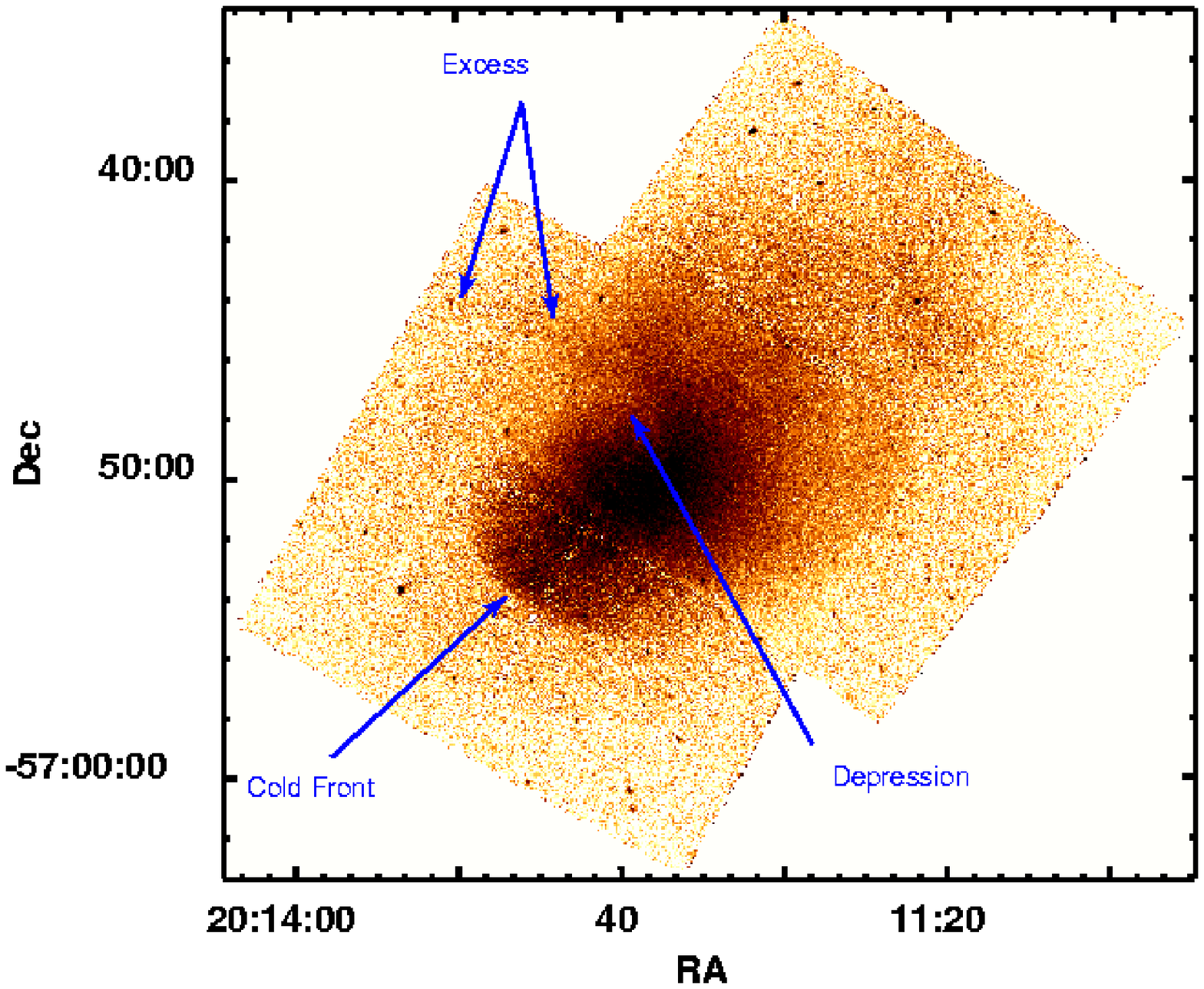}}
\caption{Count rate image in the $0.7-4$~keV band binned by $4^\second$.
The arrows indicate the prominent X-ray features: the cold front discussed by Vikhlinin et al. (2001a,b), a filamentary arc-like surface brightness excess 
extending toward the east, and a filamentary arc-like surface brightness depression extending toward the west.}   
\label{fig:image}
\end{inlinefigure}

In this letter we present the results from the joint analysis of two
 \chandra ~ observations of A3667. We identify two 
filamentary structures forming at the
border of the merging subclump: one extends toward the outskirts, 
the other toward the cluster center.
 Both structures lie well beyond the  hydrodynamically stable region
identified by Vikhlinin et al. (2001a).  
We speculate that these filaments correspond to the well-developed K-H
instability, and discuss their implications for the proposed dynamical model
of A3667.

The physical size of the structures under
  discussion is computed assuming $H_0=50$~km~s$^{-1}$~kpc$^{-1}$ 
($1$~arcsec$\approx 1.46$~kpc). Unless
specified differently, all the errors are at $90\%$ confidence level for one
interesting parameter.

\begin{inlinefigure}
\centerline{\includegraphics[width=0.95\linewidth]{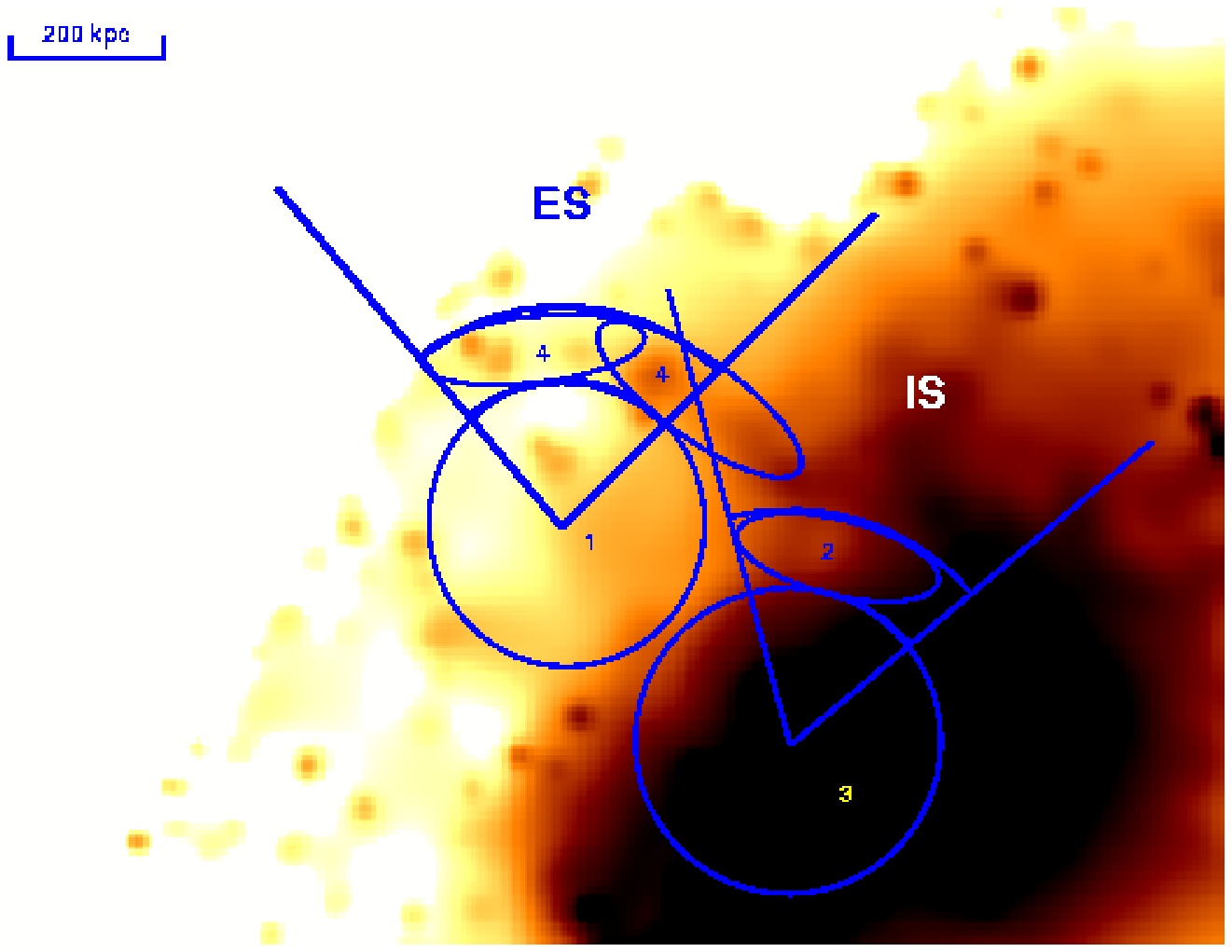}}
\caption{Adaptively smoothed \chandra ~ image. 
The four lines identify the External Sector (ES) and the Internal Sector (IS)
used to derive the surface brightness and temperature profiles reported in Fig.~\ref{fig:p_dep}. The position angles are from -15\degd ~ to 15\degd ~ and from -15\degd ~ to 40\degd ~ for IS and ES, respectively 
(angles are measured from North through East). 
Regions 4 and 2 identify the X-ray excess and depression, respectively. Regions 1 and 3 are two 
circular $r=200$~kpc regions 
centered
at the center of curvature
 of each filamentary structure. 
The actual temperatures of the regions, indicated by cardinal numbers 
1 to 4 are, reported in 
Fig.~\ref{fig:t_reg}.}
\label{fig:panda}
\end{inlinefigure}

\section{Data Reduction}\label{par:data}

A3667 was observed twice by \chandra ~ with the ACIS-I detector on
  Sept 22, 1999 and on Sept 9, 2000 for useful exposure times of 45~ks and
  50~ks, respectively. We used the data analysis procedures already
  described elsewhere (Markevitch et al.\ 2000,
  Vikhlinin et al.\ 2001a, Markevitch \& Vikhlinin 2001, and
  Mazzotta et al.\ 2001 ). We only note that, since the observations have been
  performed at different detector temperatures, the spectral response and
  background data have been generated individually for each observation and
  then appropriately combined.
  
  Spectral analysis has been performed in the 1--9~keV energy band in PI
  channels. 
  The spectral fitting has been
  performed assuming an absorption fixed at the Galactic value
  ($N_H=4.1\times 10^{20}$~cm$^{-2}$), and a fixed plasma metallicity,
  $Z=0.3\ Z_\odot$. Because of the hard energy band we used, the derived
  plasma temperatures are insensitive to the precise value of either $N_H$
  or $Z$. The temperature map has been produced with a fixed Gaussian
  smoothing of $\sigma=40^\second$ (see Vikhlinin et al.\ 2001a for the details
  of the temperature map computation).

\section{X-ray image and Temperature map}

The combined
  background subtracted and vignetting corrected 
  \chandra~ image in the 0.7-4~keV energy band is
shown in Fig.~\ref{fig:image}.
We extracted the image in the 0.7-4~keV band
to minimize the fractional contribution of the cosmic background and 
thereby to maximize the signal-to-noise ratio.  
The strongest X-ray feature present in
Fig.~\ref{fig:image} is the sharp surface brightness edge (``the cold
front'') to the South-East extensively 
discussed by Vikhlinin et al. (2001a,b). To the North of the cold front
the cluster image shows two new interesting X-ray features:

i) a filamentary X-ray excess extending toward the east to the chip boundary.
  The filament has an arc-like shape with a curvature
radius of $\approx 200$~kpc, is $\approx 300$~kpc long and $\approx 90$~kpc
wide;

ii) a filamentary X-ray depression that develops toward the west inside the
cluster center. Also this feature has an arc-like
shape with a curvature radius of $\approx 200$~kpc. It appears to be
$\approx 200$~kpc long and $\approx 80$~kpc wide;

\begin{inlinefigure}
\centerline{\includegraphics[width=0.95\linewidth]{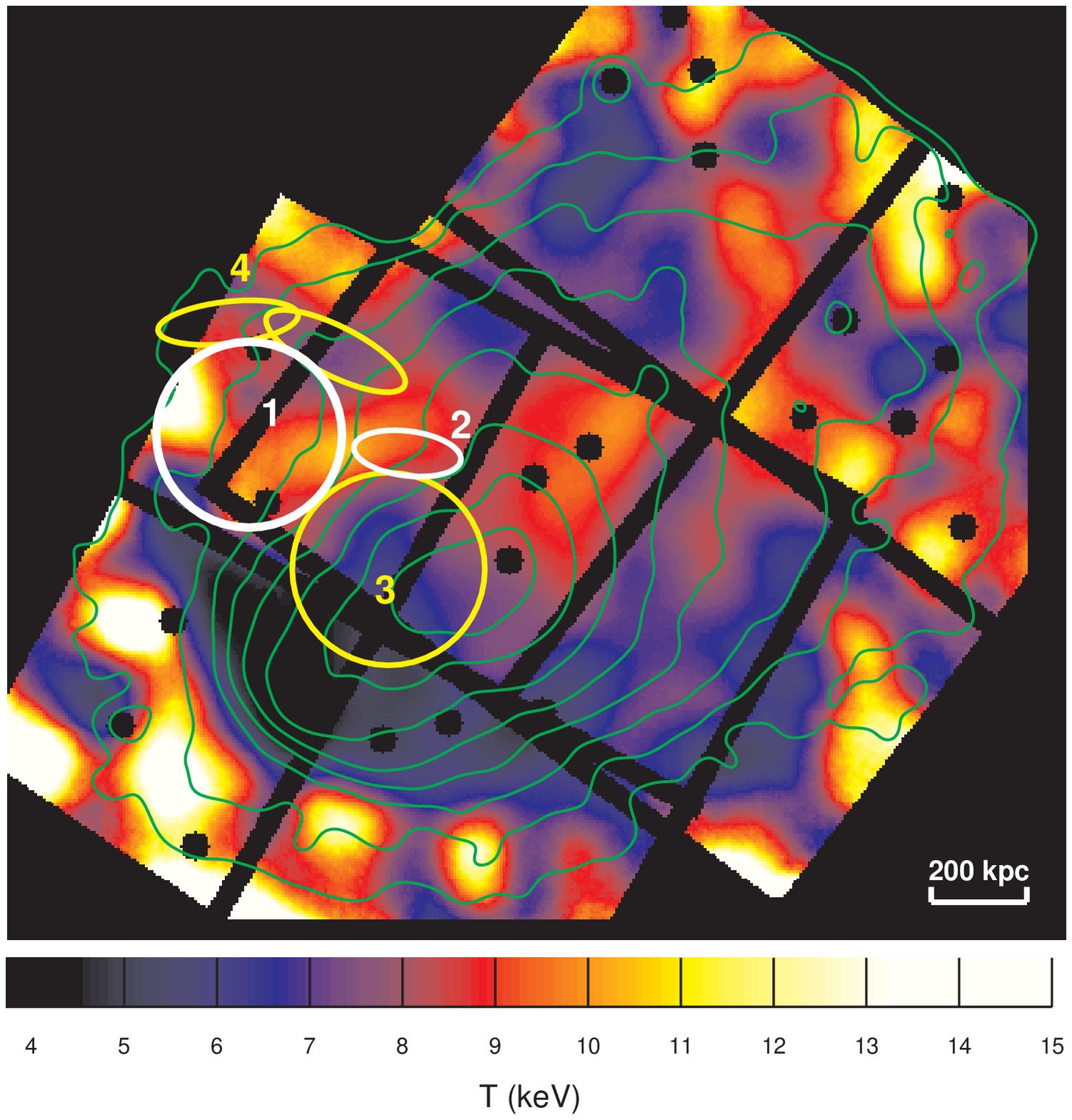}}
\caption{Temperature map  with overlaid ACIS-I X-ray
surface brightness contours (spaced by a factor of $\sqrt{2}$) in the
0.7-4~keV energy band after adaptive smoothing.   The black cut-out regions identify
  the point sources that were masked out.
The regions indicated by cardinal numbers 1 to 4 are  the same shown in 
Fig.~\ref{fig:t_reg}.}
\label{fig:tmap}
\end{inlinefigure}

\begin{inlinefigure}
\centerline{\includegraphics[width=0.95\linewidth]{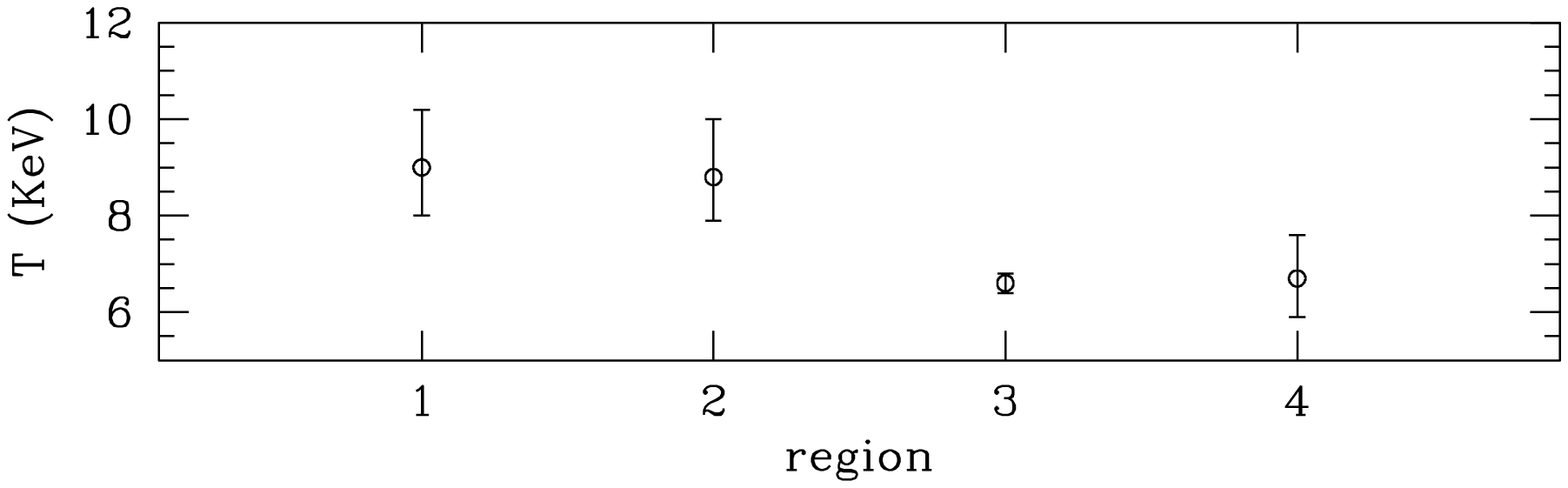}}
\caption{Projected emission-weighted temperatures in the
 corresponding regions shown in  Fig.~\ref{fig:panda} and  Fig.~\ref{fig:tmap}
($90\%$ confidence level error bars).}
\label{fig:t_reg}
\end{inlinefigure}

A close-in view of the filamentary X-ray features is shown in the 
adaptively smoothed image in Fig.~\ref{fig:panda}. We select two ellipsoidal regions indicated by the
cardinal numbers 4 and 2 to identify the X-ray excess and depression,
respectively.  To derive the surface brightness and temperature
profiles (see \S~\ref{par:prof}, below) for each feature, we identified two
sectors originating near the curvature center of the corresponding feature
and containing the feature itself.  We refer to the sector for the excess
and the depression as the External Sector (ES) and the Internal Sector (IS),
respectively.  Finally with the cardinal  numbers 1 and 3 we identify
two circular $r=200$~kpc regions centered at the origin of the ES and IS,
respectively.
 
In Fig.~\ref{fig:tmap} we report the temperature map with  overlaid X-ray
surface brightness contours. 
The temperature map 
suggests that, while the X-ray excess (region 4) corresponds 
to a local decrease in the projected temperature,
the  temperature of the X-ray depression (region 2) 
is higher than in the nearby regions. 
  To quantify the statistical significance of the features in
  the temperature map, we fitted the spectra from the four
selected regions with a single temperature model.  
The spectra from region 2 and 4 have 5155 and 3641 net photons 
plus  575 and 705 background photons, respectively.
The best-fit temperatures reported in
  Fig.~\ref{fig:t_reg} show
that while the projected temperature of the gas from the 
filamentary excess is consistent with the temperature of 
the cold subcluster, the temperature of the 
  filamentary depression is consistent with the temperature of the
ambient cluster gas.

Finally, we note that, in the
overlapping region, 
the Chandra temperature map qualitatively agrees  
with the coarser ASCA temperature map shown by 
Markevitch et al. (1999). 

\begin{inlinefigure}
\centerline{\includegraphics[width=0.95\linewidth]{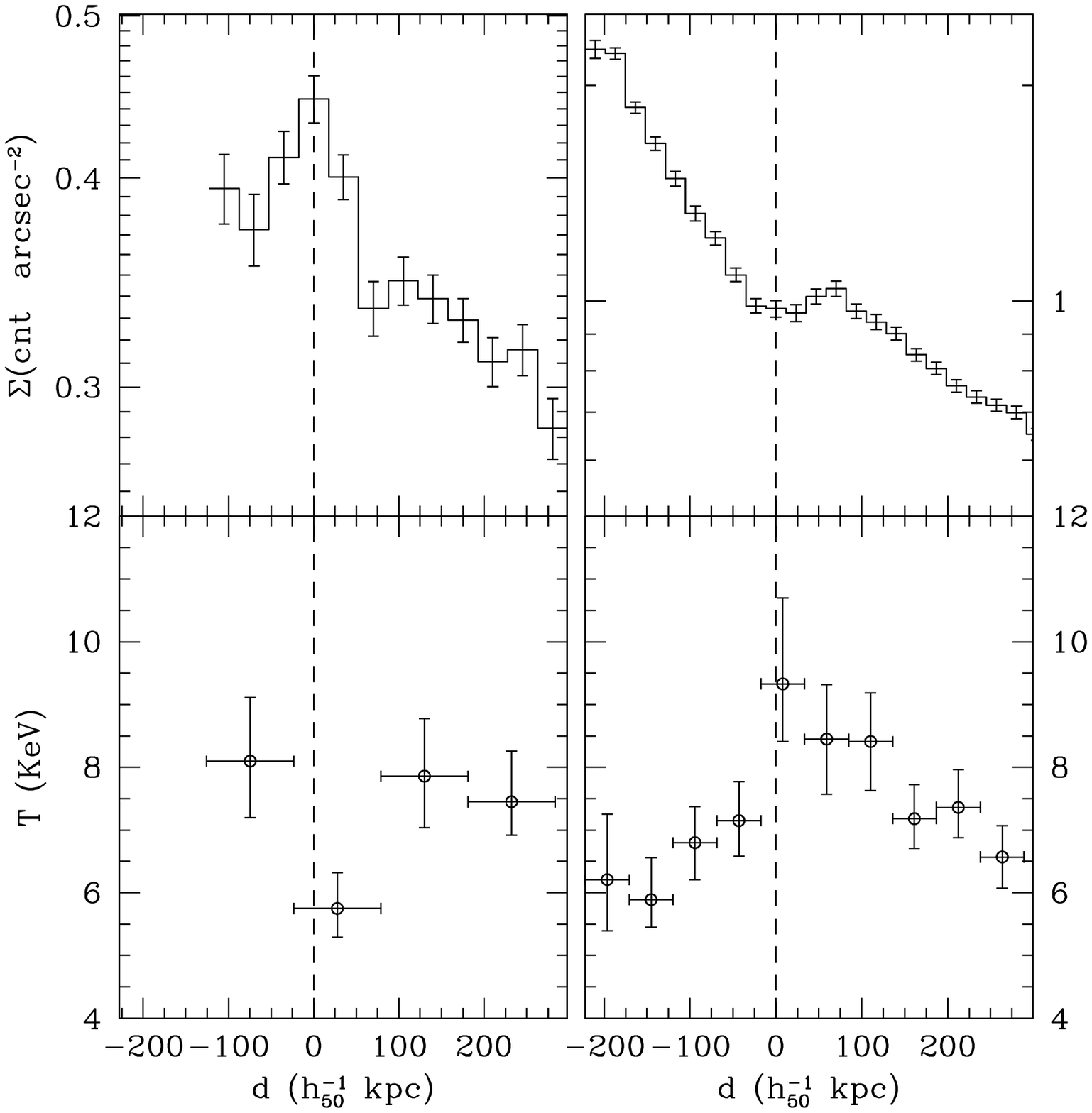}}
\caption{X-ray surface brightness (upper panels) and temperature  (lower panels)
profiles from the ES (left column) and IS (right column), 
defined in Fig.~\ref{fig:panda}. 
The dashed line in the left column indicates the relative maximum in the 
ES X-ray surface brightness profile, while 
the dashed line in the right column indicates the relative minimum in the 
IS X-ray surface brightness profile.
The x-axis    indicates the distance in kpc from the surface brightness profile stationary point 
of the corresponding sector.
Error bars are at $68\%$ confidence level.}
\label{fig:p_dep}
\end{inlinefigure}

\subsection{Surface Brightness and Temperature Profiles}\label{par:prof}

To verify the statistical significance of the observed features we extracted
both the surface brightness and the projected emission-weighted temperature
profiles.  The profiles are shown in the left and the right columns of
Fig.~\ref{fig:p_dep} for the ES and IS, respectively.  The profiles confirm
that the observed features are statistically significant.  The ES profiles
show that the filamentary excess produces a significant enhancement of the
surface brightness profile and corresponds to a decrement in the temperature
profile.  On the other hand, the IS profiles shows that the filamentary
depression produces a significant decrement of the surface brightness
profile corresponding to an increment in the temperature profile.  This
suggests that the excess is a colder, denser filamentary gas structure
embedded in the more diffuse and hotter external cluster atmosphere while
the depression is a hotter, rarefied filament of gas embedded in the denser
and colder cluster core.

\section{Discussion}\label{par:discussion}

We have presented 
evidence for the existence of two adjacent
filamentary arc-like structures in the X-ray image of A3667: an X-ray excess
extending toward the outskirts and an X-ray depression extending toward the
cluster center.  Both features are present in the soft band as well as in
the hard band. This indicates that they are produced by significant gas
density deviations with respect to the surrounding gas.  In particular, the
former indicates a denser filamentary gas structure embedded in the more
diffuse external cluster atmosphere while the latter indicates the presence
of a rarefied filament of gas embedded in the denser cluster core.  
 The
temperature analysis clearly shows that the dense external filament
is colder and the rarefied
internal filament is hotter than the ambient gas,
respectively.  We show also that, while the projected temperature of the gas
from the external filament is consistent with the temperature of the central
region, the temperature of the internal filament is consistent with the
temperature of the *ambient* external cluster gas (see Fig.~\ref{fig:t_reg}).
The observed features suggest that some dense cold gas from the cluster core
is ``stripped out'' into the hotter cluster atmosphere as well as some
rarefied hot gas from larger radii penetrates 
into the cluster core.  This observation, thus, represents good evidence for
ongoing turbulent gas mixing in the cluster atmosphere.

In  previous work, Vikhlinin et al. (2001,a,b) 
found that the gas in the  $r\approx 300$~kpc 
central, colder region of the cluster moves from North-West to
South-East into the rarefied hotter cluster atmosphere,
and the cold front is the the contact surface between
the two 
gases.  They 
determined the motion speed from the pressure jump at the
front and found the near-sonic velocity $M\equiv v/v_s \gsim 1$,
where $v_s$ is the speed of sound in the hot
ambient gas.
During the gas cloud motion, the surrounding  gas flows around the cold cloud.
In this situation, the interface between the tangentially moving gas layers
must develop both Rayleigh-Taylor (R-T) and Kelvin-Helmholtz (K-H)
instabilities (see Inogamov, 1999 for a review).

The R-T instability develops at the interface  between the two fluids when the
rarefied fluid accelerates the denser one.  If, as it is reasonable to
assume, the gas cloud moves together with its own dark matter halo 
(Vikhlinin \& Markevitch 2002, in preparation)
the gas cloud is stabilized against the R-T instability by 
gravity, so that the development of R-T
instability is quite unlikely.  In contrast the K-H instability is related
to the shearing motion at the boundary between two fluids and is expected to
develop along the lateral boundaries of the cloud. 
The wavevector  of the fastest
growing mode, $\lambda$, is parallel to the flow and its growing time $\tau$ is given by
solving the dispersion equation (see e.g. Miles 1958).
Vikhlinin et al.\ (2001b) computed the K-H instability growth time for the
flow near the cold front in A3667. 
As the only perturbations relevant for our discussion are the ones that 
grow on time scales shorter than the cluster core
passage time, $t_{cross}=L/v_{cross}$ (here $L$ is the cluster size and
$v_{cross}$ is the motion speed), we derived the ratio between 
these two times\footnote{This equation reproduces eq.(4)
  from Vikhlinin et al.\ (2001b), corrected for an algebraic error which
  resulted in overestimation of $\tau$ by a factor of $4\pi^2$. Note that
  the correct equation does not affect any results presented in Vikhlinin et
  al., and indeed, strengthens their arguments.}:
\begin{equation}
  \frac{t_{cross}}{\tau}=3.3 {L \over \lambda} \sin \varphi,
  \label{eq:1}
\end{equation}
where $\varphi$ is the angle between the perturbation and the leading edge
of the moving cloud.  

An important aspect already discussed by Vikhlinin et al.\ (2001b) is that, 
in the absence of stabilizing factors and assuming $L\approx 1$~Mpc,
 the growth time for all  perturbations on scales $\lsim 10$~kpc 
is much shorter
 than the cluster passage time  even at very small angles 
 ($\varphi \gsim 10-50$~arcmin).
 This means that  there should develop a turbulent layer which
 would smear the cold front by at least $\sim 10$~kpc at angles 
 $\varphi \gsim 10-50$~arcmin.  
 The Chandra data, however,  exclude such a smearing within 
 the $\pm30^\circ$ sector of the leading
 edge of the cold front. 
 The expected smearing appears evident only at angles 
 $\varphi > 30^\circ$ (see e.g. Fig.~\ref{fig:image}). 
 Vikhlinin et al.\ propose that the ambient magnetic field is amplified
 in a narrow
 boundary region between the two moving gas layers.
 The amplification is the result of the stretching of the magnetic field lines
 along the front by tangential plasma motions.
 The surface tension of the amplified magnetic 
 field acts to stabilize the development of the K-H instability in the 
 $\pm30$\degd ~ sector. At larger angles, however, the magnetic field 
 surface tension becomes insufficient to stabilize the front because 
 of the higher flow speed.  Thus, outside the $\pm30^\circ$
 sector, the development of the K-H instability is unaffected by the magnetic
 field, and therefore the perturbation growth time is given by
 eq.~\ref{eq:1}.
 Eq.~\ref{eq:1} gives us also two important clues to understand the presence
 of the observed
features on the side of the colder moving subcluster:

i) small scale perturbations
grow faster than the larger scale ones.  This means that small scale
perturbations develop immediately, when the interface is still a
discontinuity. Their growth, however, widens the interface, which becomes a
turbulent layer of finite width.  At this
point the perturbations on scales smaller than the evolving width of the
front are damped (they becomes unobservable as individual
structures) while perturbations on a larger scale start to
grow (thus they may appear as distinct  structures) 
(see e.g. Esch 1957). As this process
continues, it is expected that the remaining observable wavelengths are the
ones with a growth time comparable to the cloud crossing time, namely
$t_{cross}/\tau\sim 1-10$;

ii) for a fixed wavelength the growth time is shorter at larger angles
$\varphi$ reaching its minimum at $\varphi=90$\degd .  This is a direct
consequence of the fact that the speed of the external fluid increases with
$\varphi$ being maximum at $\varphi=90$\degd .
  
In this picture the most natural explanation for the newly discovered
filamentary structures in A3667 is that they are the result
of the development of K-H instabilities.  
The structure lies, in fact, at $\varphi \approx 90$\degd ~
which,  because of point  
ii) above, is indeed a privileged point for the growth
of the perturbations.
Furthermore, if we assume that $\lambda=300$~kpc and
$L=1$~Mpc, the growing time for this perturbation 
is such that $t_{cross}/\tau\approx 10$. This means
that the perturbation is just now entering the strong non-linear regime,
 consistently with what we observe.

We notice that  the observed instability act to:
i) deposit low density hot gas right in the cluster center; 
ii) remove denser colder gas from the cluster center. 
If efficient, these two processes 
may contribute to inhibit (or, if already present, to destroy)
the development of a central cooling flow.
We calculated the  mass rate of 
hot gas which is being deposited right 
in  the cluster center.
For the scope of this letter we use a very simple  model and  
we assume that the external hot gas flows toward the center 
in a cylinder-like filament with a speed $v$ close to the speed
of the subclump ($v\approx M v_c\approx 1400{\rm km~s}^{-1}$).
Moreover, we assume that the gas density $n_0$ in the filament is equal to 
the density outside the front and that the radius
of the filament is $r=40$~kpc. We Find:
\begin{equation}
\dot M=108 
\Big(\frac{r}{40\rm{kpc}}\Big)^2\Big(\frac{v}{1400\rm{km/s}}\Big)\Big(\frac{n_0}{8\times 10^{-4}/\rm{cm}^{3}}\Big)\frac{\rm{M}_\odot}{\rm{yr}}
  \label{eq:2}
\end{equation}
From Eq.~\ref{eq:2} we see that the  
instability  transports hot gas in the cluster center at a rate which is comparable to the  
mass deposition rate of many cooling flow clusters.
 It is clear, than, that such a mechanism may play an important role 
in the development and/or evolution of a central cooling flow.

\section{Conclusion}

We presented the combination of two \chandra ~ observations of the central
region of A3667.  We showed that the cluster
hosts two arc-like adjacent filamentary structures: one,
extending from the colder subcluster toward the cluster
outskirts, appears as a dense structure embedded in the less dense cluster
atmosphere, and the other, extending inside the
  subcluster, appears as a rarefied structure
embedded in the denser cluster core.

We suggest that the observed features represent the first evidence for the
development of a large scale hydrodynamic instability in the cluster
atmosphere.  This interpretation appears to be consistent with the
previous cluster dynamic interpretation proposed by Vikhlinin et al.
(2001a,b).   

The discovery of this instability represents an important step toward
a full understanding of the physics of mergers.
In particular it shows that, although
the cold front prevents the gases of the merging objects to mix along the
direction of motion, strong turbulent mixing processes, on scales comparable
to the size of the merging subclump, may occur at large angles to the
direction of motion. 
This may favor the deposition of a non-negligible quantity of
thermal energy right in the 
cluster center  with important consequences for the development and/or 
evolution  of a central  cooling flow.

\acknowledgments
We thank M. Markevitch and W. Forman for useful comments and suggestions.
P.M. acknowledges an ESA fellowship and thanks the Center for
Astrophysics for its hospitality. 
Support for this study was provided
by NASA contract NAS8-39073, grant NAG 5-9217, and by the Smithsonian
Institution.

\end{document}